\documentclass{webofc}
\usepackage[varg]{txfonts}   % Web of Conferences font
\usepackage{hyperref}
\usepackage{url}
\usepackage{overpic} 
\hypersetup{colorlinks=true,citecolor=blue,urlcolor=blue,linkcolor=blue}
\begin{document}
\title{Search for $\boldsymbol{\pi_1(1600)}$ in a three-pion system at GlueX}

\makeatletter
\renewcommand{\the@institutecpt}{}
\makeatother

\author{
\firstname{Ilia}~\lastname{Belov}\thanks{\email{Ilia.Belov@ruhr-uni-bochum.de}} \and \firstname{Farah}~\lastname{Afzal}
\firstname{}\lastname{for the GlueX Collaboration}
}

\institute{Ruhr-Universität Bochum, Universitätsstraße 150, 44801 Bochum, Germany}

\abstract{The GlueX experiment in Hall D at Jefferson Lab enables studies of the light meson spectrum in $\gamma p$ interactions with a linearly polarized photon beam. GlueX aims in particular to search for hybrid mesons that have exotic quantum numbers and therefore cannot be classified as conventional hadrons. We present the search for the $\pi_1(1600)$ meson by means of a partial-wave analysis of the $\pi^{+}\pi^{-}\pi^{-}$ system produced off the $\Delta^{++}$-baryon. Data with the selected three-pion final state are fitted in bins of $m_{3\pi}$ as coherent sums of partial-wave amplitudes defined in the reflectivity basis. The properties of $a_{2}^{-}(1320)$ production are investigated through the extracted signal in the $\rho\,\pi^{-}$ $D$-waves. In this analysis, the overall $m_{3\pi}$ lineshape will be extracted for each of the model contributions. The main interest lies in establishing the existence of a resonant $1^{-+}$ contribution in the $\rho\,\pi^{-}$ $P$-wave configuration.
}
\maketitle

\section{Introduction}
Experimental observation of hybrid mesons, quark-antiquark states with the contribution of an excited gluonic field, is a fundamental test of QCD, since such mesons are not accommodated within the spectrum of conventional quark model states, but are explicitly predicted by Lattice  QCD~\cite{Dudek:2013yja}. The most famous spin-exotic candidate is the $\pi_1(1600)$, for which evidence has been reported by different experiments in the $\eta\,\pi$, $\eta^{\prime}\,\pi$, $\rho\,\pi$, $b_{1}\,\pi$, and other decay modes~\cite{Ketzer:2019wmd}.

The $\pi_{1}(1600)$ observations rely primarily on  measurements in diffractive production with a pion beam. The first observations of $\pi_{1}(1600)$ in the $\pi^{+}\pi^{-}\pi^{-}$ final state were made independently by two experiments in the 1990s, E852 at BNL~\cite{Adams:1998mbq} and VES at IHEP~\cite{Zaitsev:2000rc}. VES obtained a consistent description of the $\pi_{1}(1600)$ signal only from a combined fit to a few decay channels, whereas measurement in the three-pion data alone was deemed unreliable~\cite{Amelin:2005ry}. E852 later disproved their observation with reanalysis of a larger $\pi^{+}\pi^{-}\pi^{-}$ dataset, recalling to the incomplete wave set in the original study and related leakage~\cite{Dzierba:2005jg}. In recent years, several COMPASS measurements reported the extraction of a $1^{-+}$ resonance in the $\pi^{+}\pi^{-}\pi^{-}$ data from a pion-beam diffractive production~\cite{Adolf:2015gxz}. Evidence was also reported in the counterpart $\pi^{-}\pi^{0}\pi^{0}$ mode~\cite{Nerling:2012ei}. A thorough comparison of the COMPASS results with each other and with previous measurements revealed that differences are mainly caused by selected momentum transfer range and applied set of partial waves~\cite{Alexeev:2021ogp}.

Photoproduction experiments provide complementary data in regards of the  production mechanism. Meanwhile, the absence of a $\pi_{1}(1600)$ signal in the $\pi^{+}\pi^{+}\pi^{-}$ mode at CLAS~\cite{Nozar:2008zko}, calls for an independent analysis of the three-pion mode in photoproduction. GlueX data are especially relevant thanks to the recently acquired high-precision dataset of 125~pb$^{-1}$ collected in 2017$-$2018~(GlueX-I data).

The GlueX experiment in Hall D at Jefferson Lab explores the spectrum of both conventional and hybrid mesons through advanced searches with partial-wave analysis techniques~\cite{Albrecht:2024qdh,Schertz:2023hnk,Scheuer:2024sad}. The experimental facility operates a tagged linearly polarized photon beam in the 8.2$-$8.8~GeV energy range incident on a liquid hydrogen fixed target. The detector setup consists of a spectrometer with nearly full solid angle coverage, which has excellent capabilities for reconstruction of charged particle tracks, reconstruction of electromagnetic showers, and charged particle identification. Detailed descriptions of the GlueX beamline and detector are published in~\cite{GlueX:2020idb}.

\begin{figure}[t]
\centering
\begin{overpic}[width=0.32\linewidth]{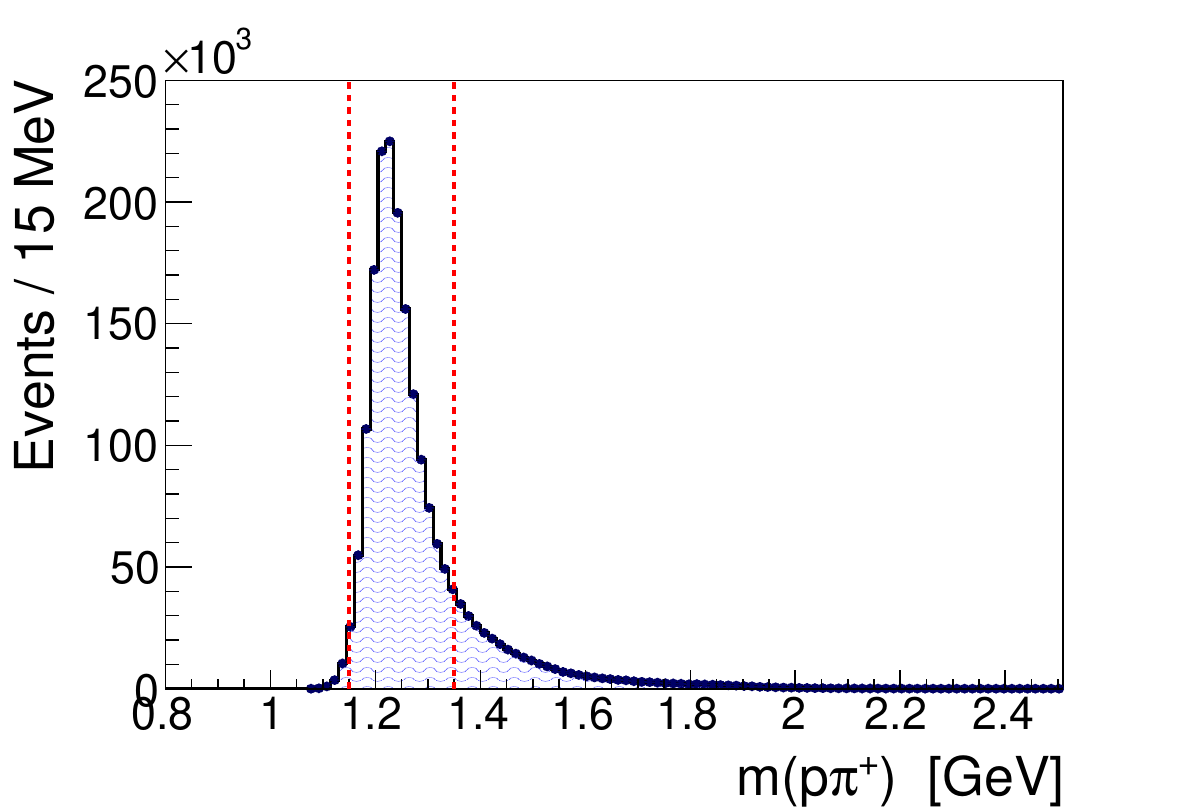} 
\put(60,45){\includegraphics[scale=0.03]{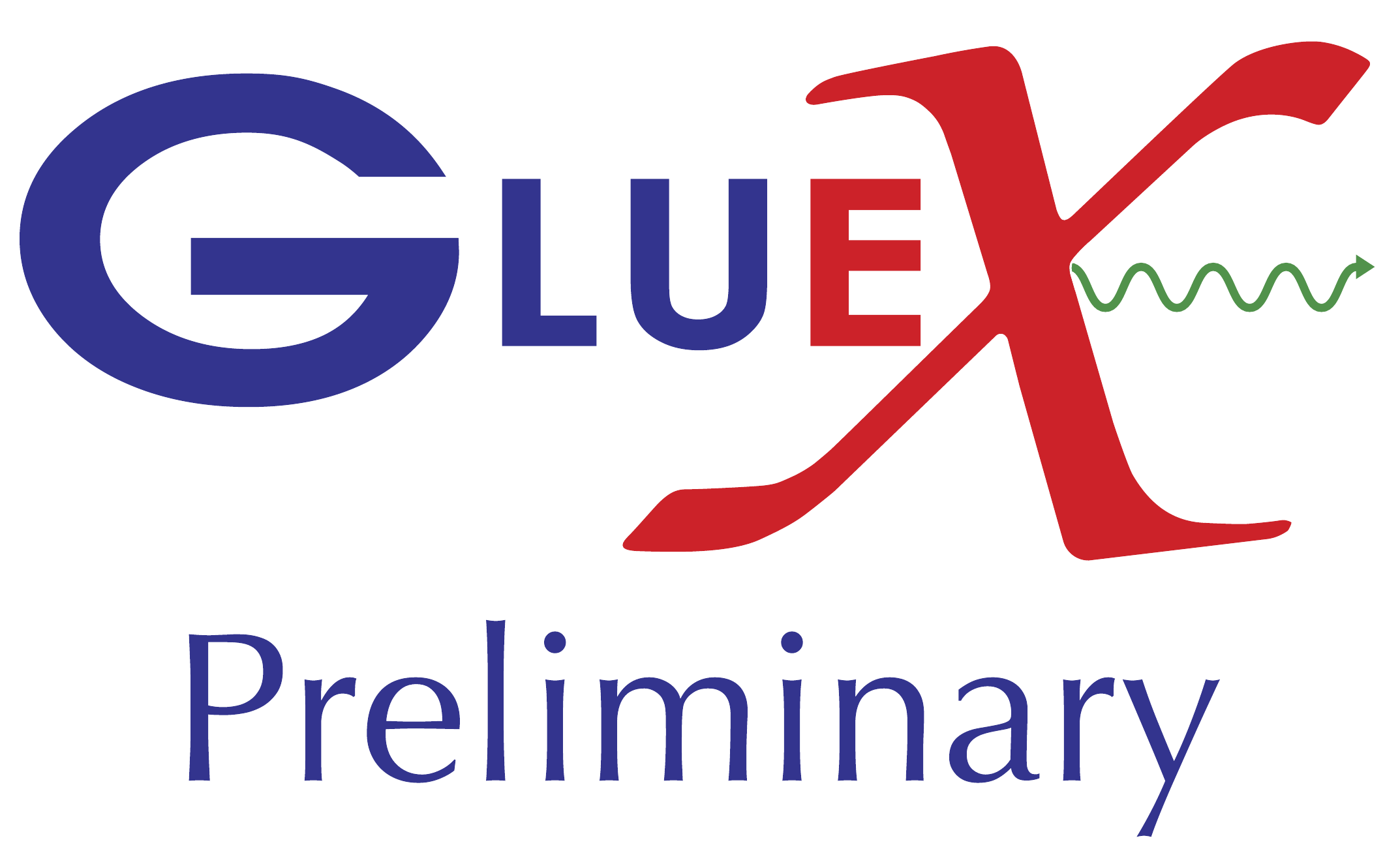}}
\end{overpic}
\includegraphics[width=0.32\linewidth]{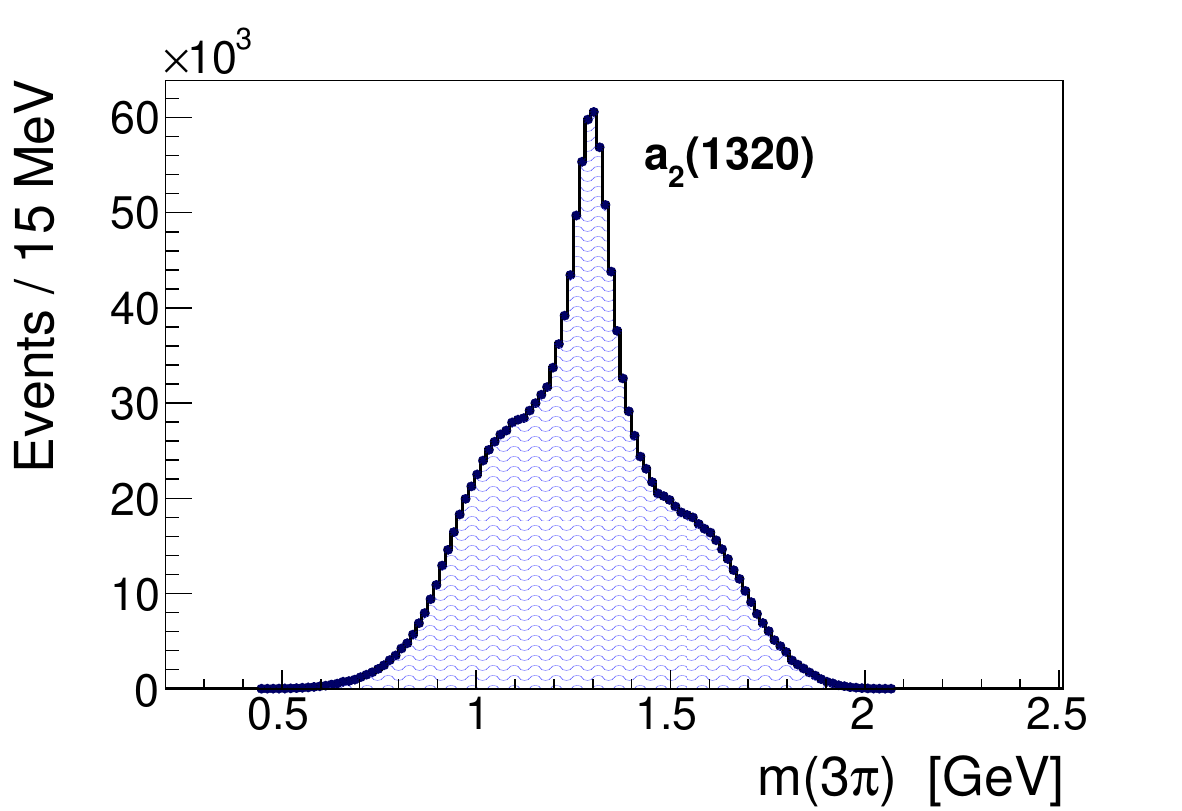} 
\includegraphics[width=0.32\linewidth]{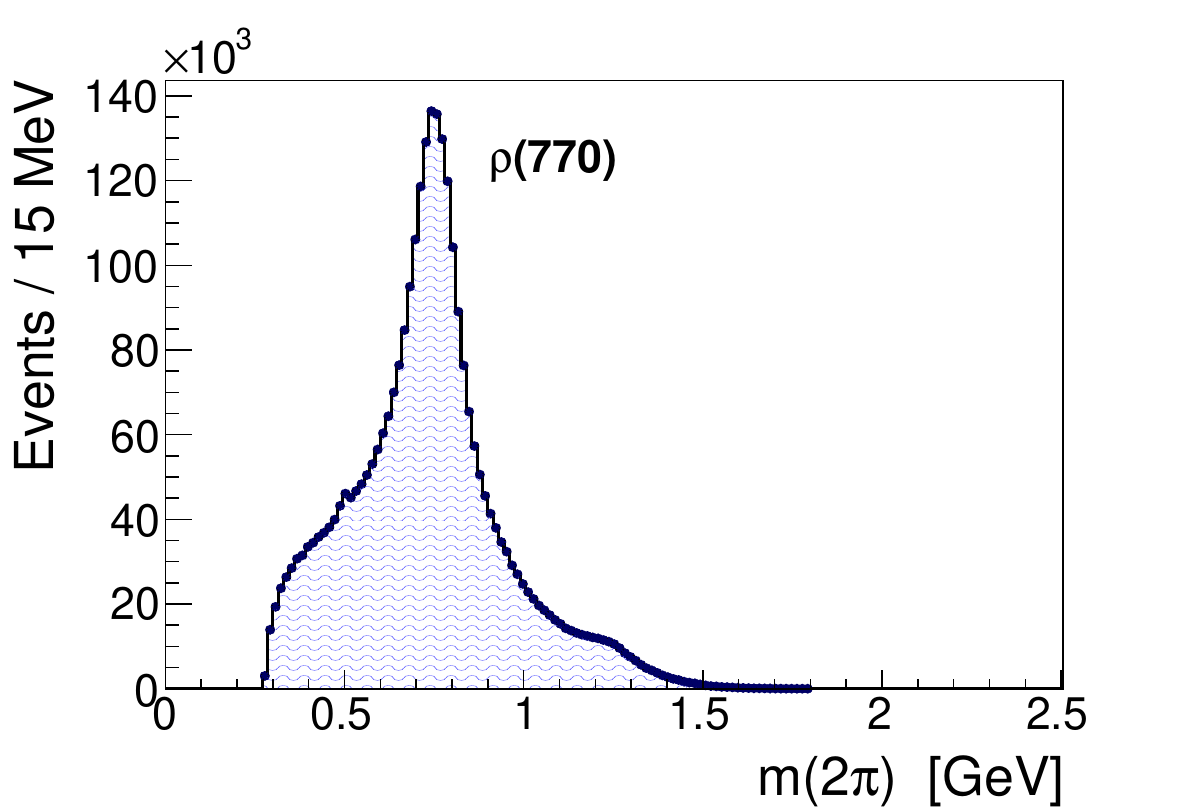} 
\caption{Selected data in the $-t\leq 0.25~\text{GeV}^2$ subregion. Shown are the invariant masses of $p\pi^{+}$~(baryon system), $\pi^{+}\pi^{-}\pi^{-}$~(meson system) and $\pi^{+}\pi^{-}$~(isobar in the meson system).}
\label{fig:data-selected}
\end{figure}

To select the $\pi^{+}\pi^{-}\pi^{-}$ final states, we retain $\gamma p\to \pi^{+}\pi^{-}\pi^{-}\Delta^{++}(\to p\pi^{+})$ exclusively reconstructed events from the GlueX-I dataset.
Using a sideband subtraction technique, we subtract the background from initial state photons mistakenly associated with the primary interaction. By sorting $\pi^{+}$'s over their momentum magnitude and imposing a cut on the fast--slow momentum difference, we are able to kinematically distinguish the two final-state $\pi^{+}$ mesons and assign them to the meson ($\pi^{+}\pi^{-}\pi^{-}$) and baryon ($p\pi^{+}$) systems. The successful separation is supported by Monte Carlo studies, demonstrating a purity of 99\% in simulated data. Significant background arises from other combinations of final state particles in the meson and baryon systems. We suppress this type of background with the help of  $p_{z}(\pi^{-}_{\text{slow}})\geq 700~\text{MeV}$ and $m_{4\pi} \geq 1800~\text{MeV}$ selections. Finally, we apply a mass-window cut around the $\Delta^{++}$ peak, $1150~\text{MeV}\leq m(p\pi^{+})\leq 1350~\text{MeV}$. Figure~\ref{fig:data-selected} shows selected data for the $-t\leq 0.25~\text{GeV}^2$ subregion, where we perform the initial fits to 1.48~M out of the total 4.33~M events.

\section{Fit with a partial-wave model independent of $\boldsymbol{m_{3\pi}}$}
A first partial-wave model for the $\eta\,\pi$ photoproduction was formulated in~\cite{Mathieu:2019fts}. In GlueX, this model was not only applied to the $\eta\,\pi$ channel~\cite{Albrecht:2024qdh}, but also adapted and applied to the vector-pseudoscalar $\omega\,\pi$ channel~\cite{Schertz:2023hnk,Scheuer:2024sad}. We in turn adapt and apply this model to describe the $\xi\,\pi$ pair, assuming a three-pion decay proceeds via intermediate isobars~$\xi(\to\pi^{+}\pi^{-})$ of various~$J^{P}$. We analyze data in the $-t\leq 0.25~\text{GeV}^2$ subregion, in bins of the invariant $3\pi$ mass. The measured intensity is fitted with  
\begin{equation}
{\cal I}_{\textrm{tot}} = {\cal I}(\Phi,\Omega,\Omega_{H},m_{2\pi})\cdot \textrm{BW}(m_{\Delta^{++}}) + {\cal I}_{\textrm{uniform}},
\end{equation}
where the baryon system angles are integrated out, the two nucleons' spin orientations are summed over, ${\cal I}(\Phi,\Omega,\Omega_{H},m_{2\pi}) \equiv {\cal I}(\Phi,\tau)$ is the signal intensity with partial waves, and ${\cal I}_{\textrm{uniform}}$ is the background of uniformly distributed phase space events. Coherent sums are constructed in the reflectivity basis, including both positive and negative ($\varepsilon = \pm$) reflectivities,
\begin{multline}
{\cal I}(\Phi,\tau) \sim \Bigg[ 
\left(1-P_{\gamma}\right)\bigg|\sum_{i,m}\Big[{\cal T}_i\Big]_{m}^{(-)} \text{Im}\big[Z^{(i)}_{m}(\Phi,\tau)\big]\bigg|^{2} + \left(1-P_{\gamma}\right)\bigg|\sum_{i,m}\Big[{\cal T}_i\Big]_{m}^{(+)} \text{Re}\big[Z^{(i)}_{m}(\Phi,\tau)\big]\bigg|^{2} \\
\label{coherent-sum}
+ \left(1+P_{\gamma}\right)\bigg|\sum_{i,m}\Big[{\cal T}_i\Big]_{m}^{(+)} \text{Im}\big[Z^{(i)}_{m}(\Phi,\tau)\big]\bigg|^{2} + \left(1+P_{\gamma}\right)\bigg|\sum_{i,m}\Big[{\cal T}_i\Big]_{m}^{(-)} \text{Re}\big[Z^{(i)}_{m}(\Phi,\tau)\big]\bigg|^{2} \Bigg], 
\end{multline}
where $P_{\gamma}$ is the photon beam polarization fraction, the index~$i$ spans over different $J^{P}L$ configurations written with a shortcut ${\cal T}$, and the index~$m$ runs over all the spin projections from $-J$ to $J$. The decay amplitudes $Z^{(i)}_{m}(\Phi,\tau)$ capture dependence on the angle $\Phi$ between the beam polarization vector and the production plane,  angles in the $3\pi$ system ($\Omega$ -- angles of the $\xi$ momentum in the $3\pi$ rest frame, and $\Omega_{H}$ -- angles of $\pi^{-}$ momentum in the $\xi$ rest frame), and the isobar mass $m_{2\pi}$. Given the two possible amplitude signatures, $\tau_{a}$ and $\tau_{b}$, resulting from the two kinematically indistinguishable $\pi^{-}$ in our data, we apply Bose-symmetrization as
$Z^{(i)}_{m}(\Phi,\tau) = \frac{1}{\sqrt{2}}\left[Z^{(i)}_{m}(\Phi,\tau_{a}) + Z^{(i)}_{m}(\Phi,\tau_{b})\right]$.

\begin{figure}[t]
\centering
\begin{overpic}[width=0.325\linewidth]{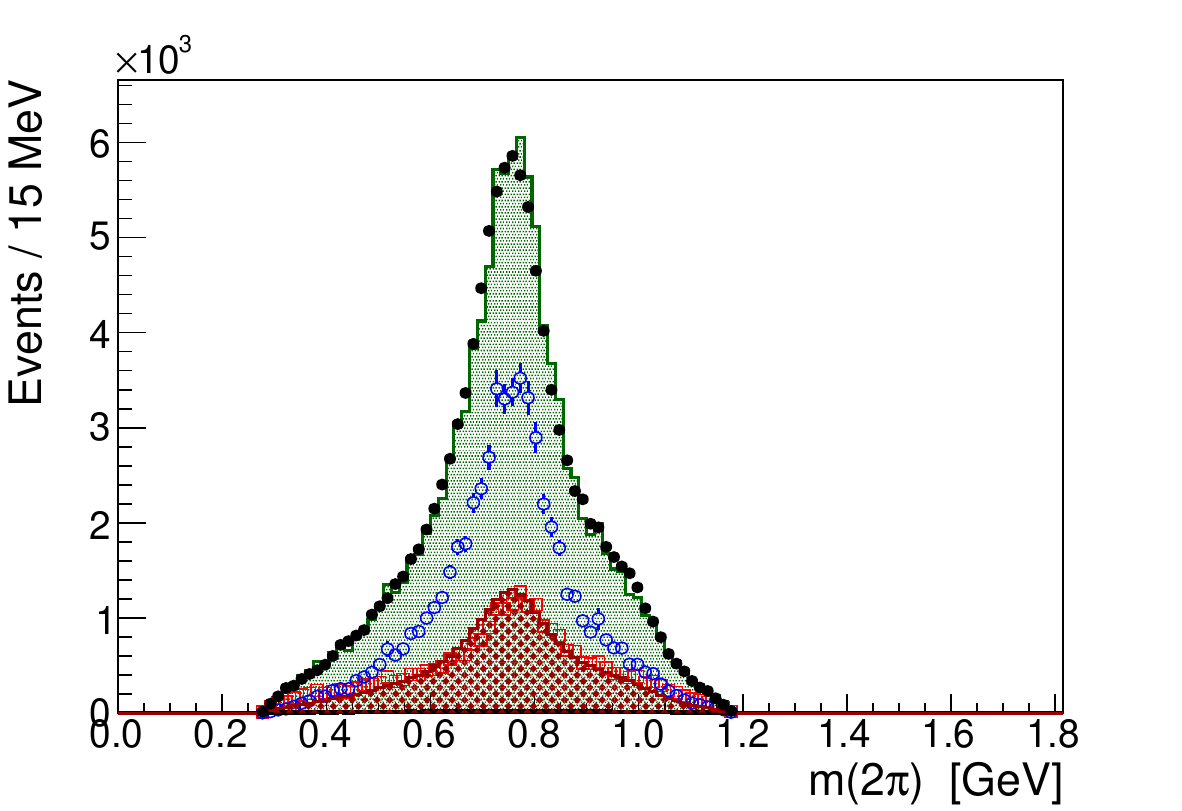 } 
\put(62,45){\includegraphics[scale=0.03]{Pics/Logo_GlueX_preliminary}}
\end{overpic}
\hfill
\includegraphics[width=0.325\linewidth]{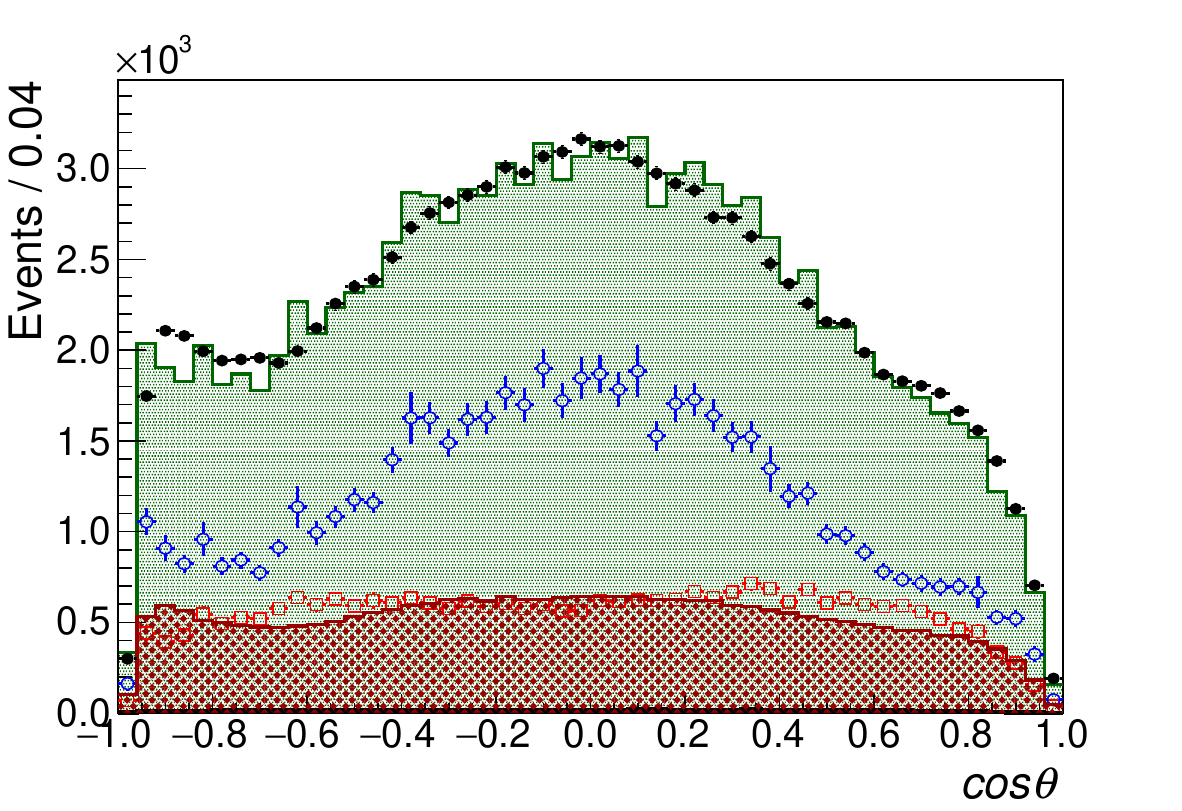 } 
\hfill
\includegraphics[width=0.325\linewidth]{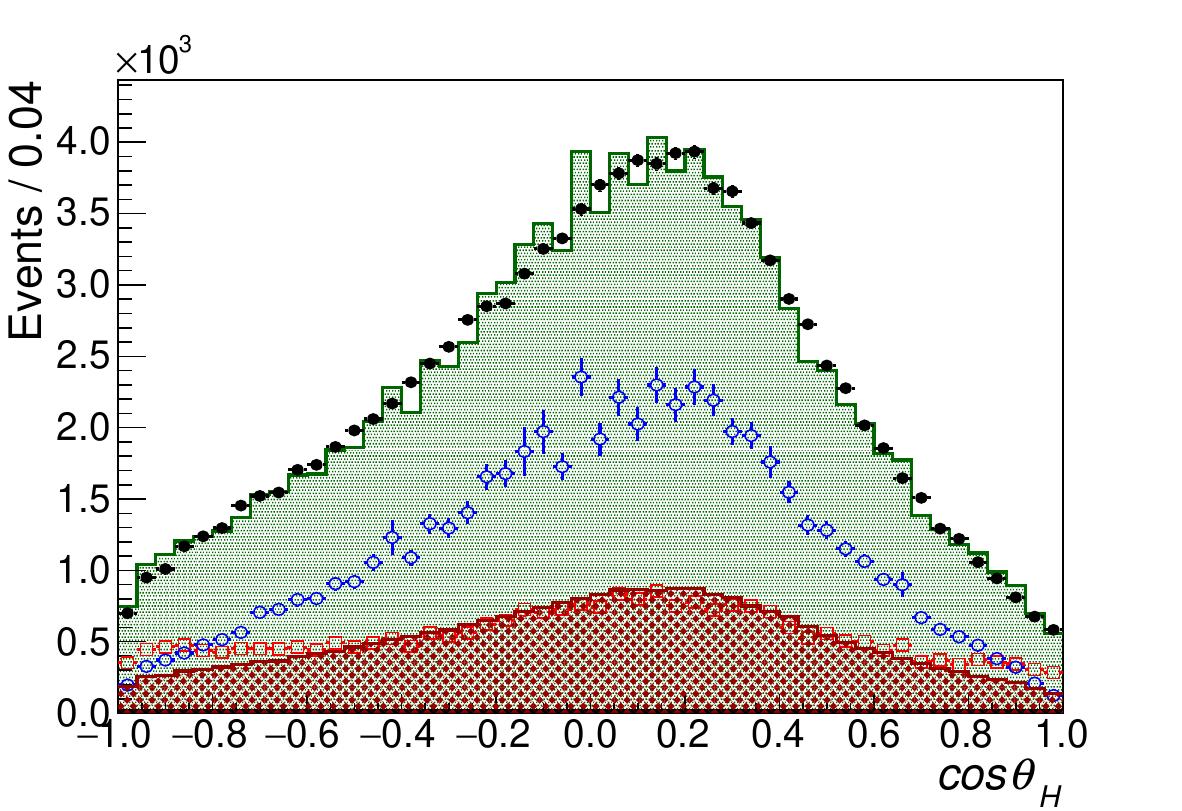 } 
\vfill
\includegraphics[width=0.325\linewidth]{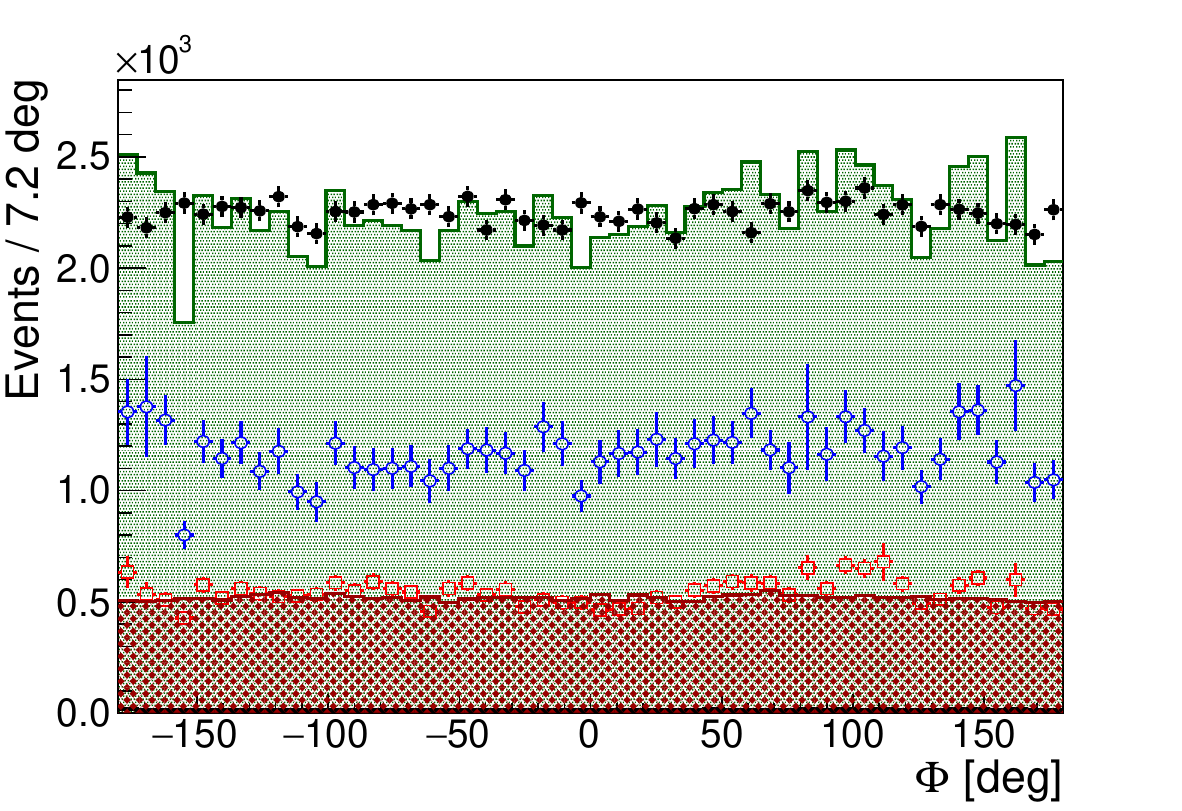 } 
\hfill
\includegraphics[width=0.325\linewidth]{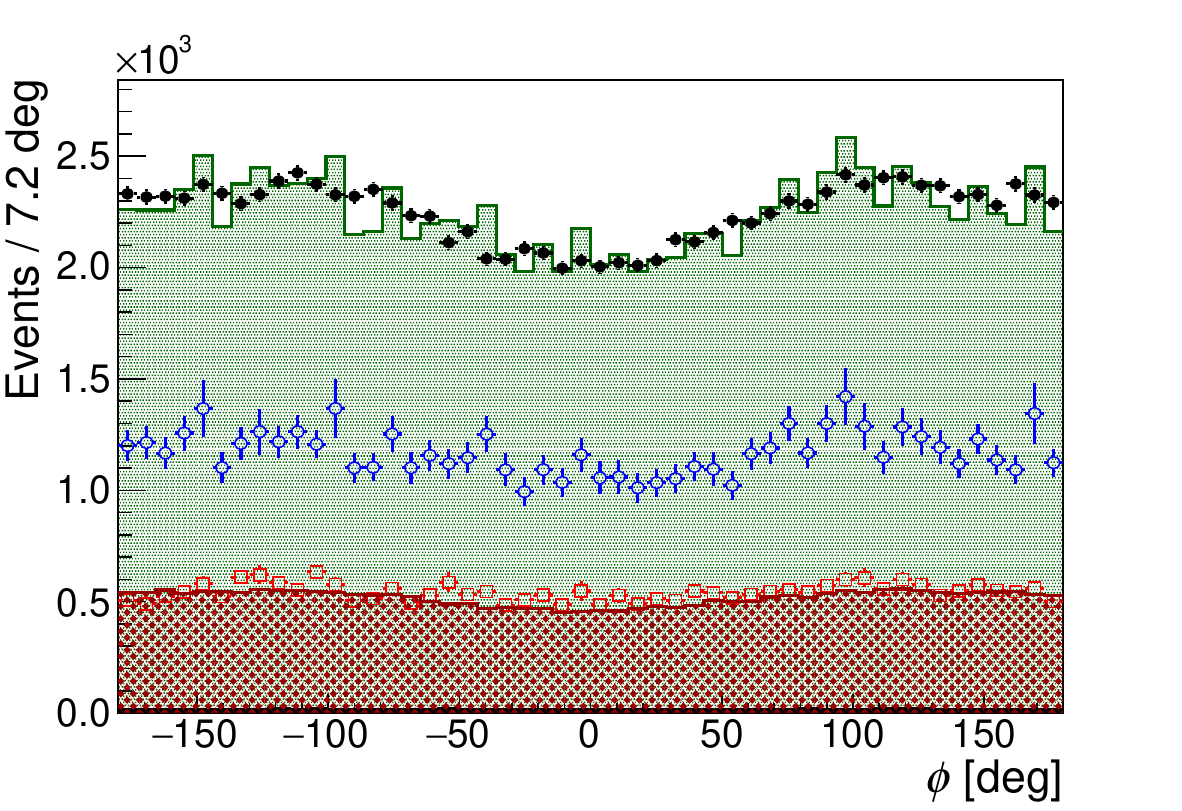 } 
\hfill
\includegraphics[width=0.325\linewidth]{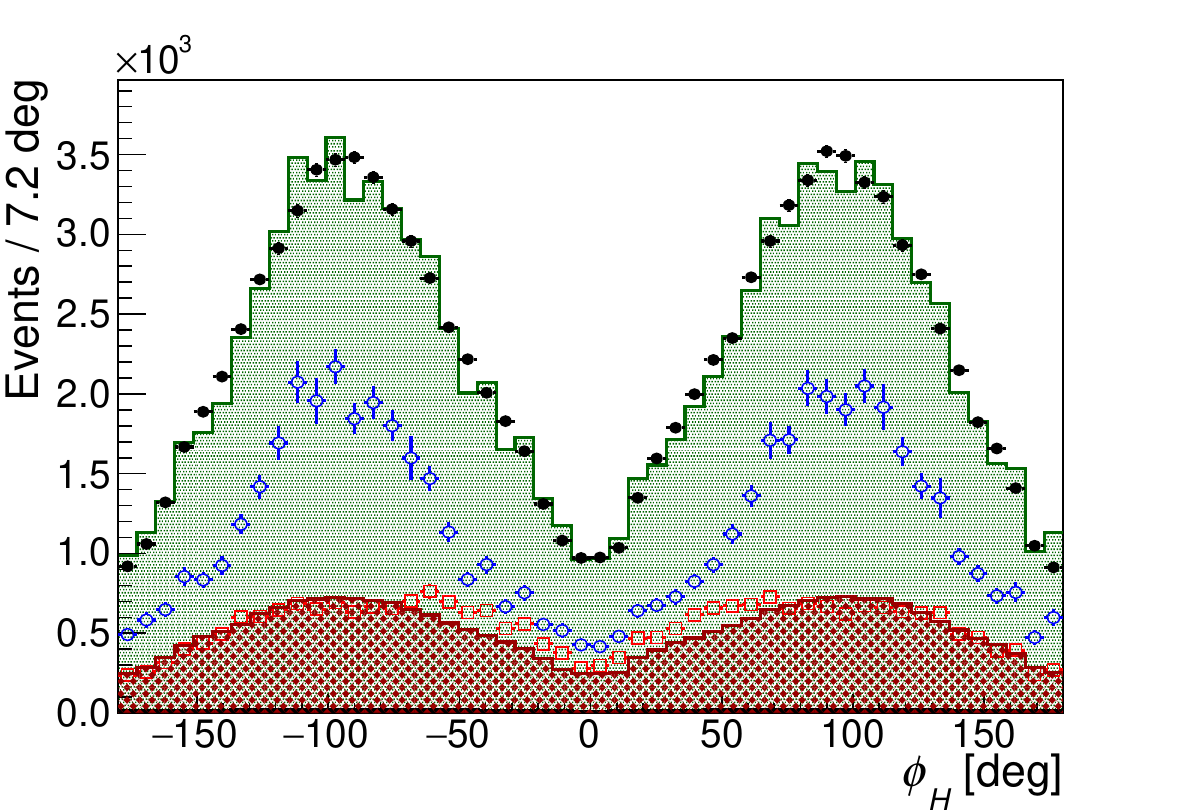 } 
\caption{Preliminary mass spectrum and angular distributions of the partial-wave fit in the $1300\,\text{MeV}\:{\leq}\:m_{3\pi}\:{\leq}\:1325\,\text{MeV}$ bin: data points~(black), total fit~(green), accidental photon background~(brown), total contributions of positive reflectivity~(red) and negative reflectivity~(blue).}
\label{fig:fit-projections}
\vspace*{-2.3ex}
\end{figure}

We extract the partial-wave contributions, which are the complex-valued fit parameters $\Big[{\cal T}_i\Big]_{m}^{(\varepsilon)}$ in Eq.~\eqref{coherent-sum}, from extended unbinned likelihood fits to data. Fits are performed with the \texttt{AmpTools} library~\cite{Shepherd:2024}, which employs the \texttt{Minuit} package for numerical minimization. In this paper, we discuss fits with a pool of 36 partial waves, 
\begin{equation}
\label{waveset}
\left\lbrace \left[\pi\pi\right]_{S}\pi\:\big|\:0^{-}S,~1^{+}P\right\rbrace,\qquad
\left\lbrace \rho\,\pi~~\big|\:0^{-}P,~1^{+}S,~2^{+}D\right\rbrace,\qquad 
\left\lbrace f_{2}\,\pi\:\big|\:2^{-}S\right\rbrace,
\end{equation}
where the notation $[\xi\pi]\,{:}\,J^{P}L$ is used, each with all possible values of $\varepsilon$ and $m$, since we found those to be satisfying sets of waves describing the data well in the entire $m_{3\pi}$ range. The $m_{2\pi}$ lineshape of each of the three isobars is modelled by an explicit factor in $Z^{(i)}_{m}(\Phi,\tau)$. For the $\rho$ and $f_{2}$ isobars, we make use of relativistic Breit-Wigner lineshapes with both mass and width fixed to the PDG values. For the  $[\pi\pi]_{S}$ isobar, we follow I.~Kachaev~\cite{Kachaev:2023tsx} and make use of an effective formula providing a broad and smooth lineshape with a small phase motion, excluding $f_{0}(980)$, as needed to describe the $m_{2\pi}$ mass in the region of low $m_{3\pi}$ in our data. Finally, we fit our data with the wave set of Eq.~\eqref{waveset} in 25~MeV bins across the full range of $m_{3\pi}$ shown in Figure~\ref{fig:data-selected}  and observe fairly good fits in every $m_{3\pi}$ bin.

\section{Results for the $\boldsymbol{a_{2}(1320)}$ production}
Investigating the production mechanism of $a_{2}(1320)$ is a crucial step towards developing models for photoproduction at GlueX energies that can further aid in the search for hybrid mesons. Moreover, the $a_2(1320)$ signal serves as an important benchmark for our analysis, since the fit results can be directly compared to those of partial-wave analysis of the $\eta\,\pi^{-}$ system at GlueX, where a contribution from $a_{2}(1320)$ is significant~\cite{Albrecht:2024qdh}. For these reasons, we present fit projections for $m_{3\pi}$ in the vicinity of 1320~MeV and focus on the comparison of the $2^{+}$ waves.

\begin{figure}[ht]
\centering
\begin{minipage}{0.42\linewidth}
\centering
\begin{overpic}[width=1\linewidth]{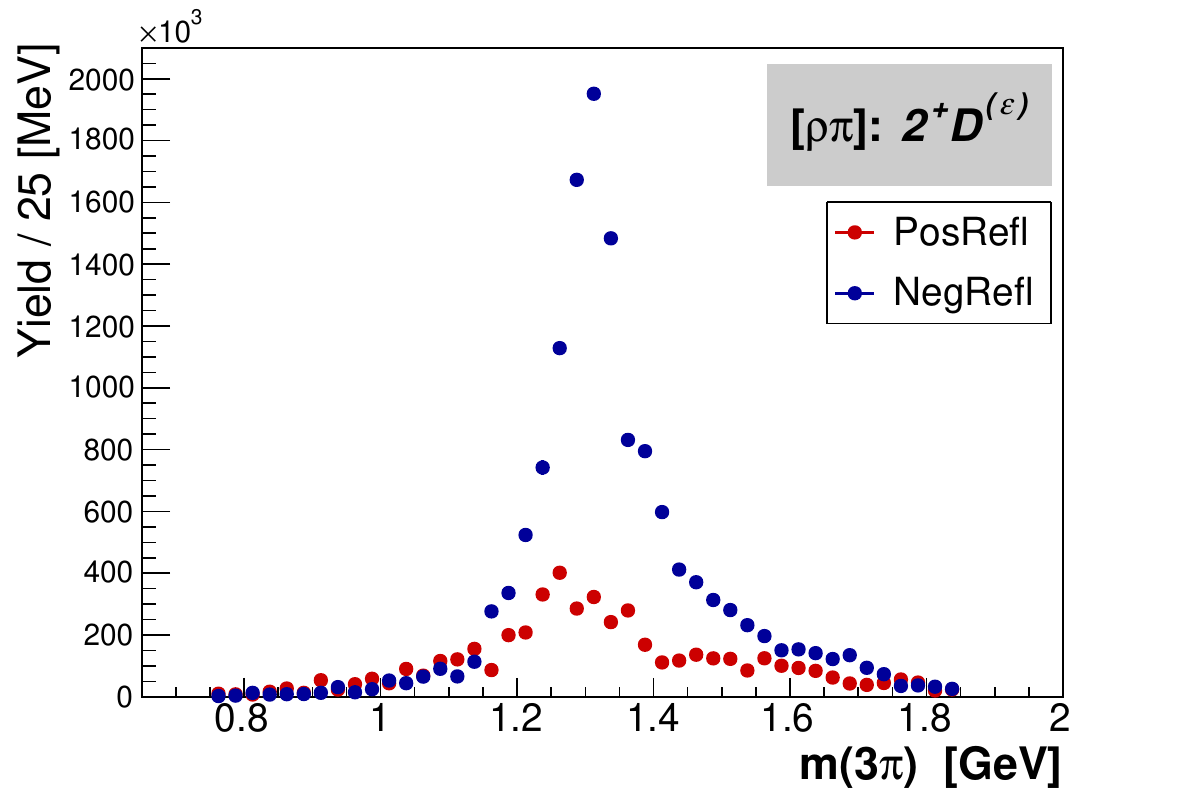} 
\put(15,50){\includegraphics[scale=0.03]{Pics/Logo_GlueX_preliminary}}
\end{overpic}
\end{minipage}
\begin{minipage}{0.46\linewidth}
\centering
\vspace{1ex}
\includegraphics[width=1\linewidth]{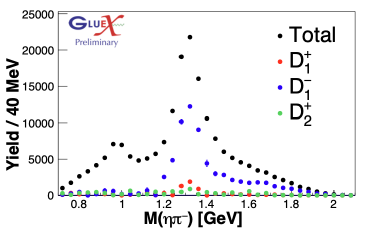} 
\end{minipage}
\caption{Preliminary results of a partial-wave analysis for the dominant $2^{+}D$-wave intensities exhibiting the $a_{2}(1320)$ signal. Left panel: $\pi^{+}\pi^{-}\pi^{-}$ mode; right panel: $\eta\,\pi^{-}$ mode, figure taken from~\cite{Albrecht:2024qdh}.}
\label{fig:a2-comparison}
\end{figure}

Figure~\ref{fig:fit-projections} shows good agreement of the fit with data in the fit projections in the vicinity of 1320~MeV, displaying the contributions from total coherent sums with both reflectivities. One can also see the enhanced negative reflectivity contribution. This result implies the prevalence of unnatural $t$-channel exchange amplitudes in $a_{2}(1320)$ production off the $\Delta^{++}$, such as amplitudes of a single pion exchange. The left panel of Figure~\ref{fig:a2-comparison} presents a clear peak for the $a_{2}(1320)$ reconstructed in $3\pi$ invariant mass as a $[\rho\pi]\,{:}\,2^{+}D$ coherent sum. Given the prominent $a_2(1320)$ and $\rho(770)$ signals in data, the $[\rho\pi]\,{:}\,2^{+}D$ waves should be dominating the partial-wave intensity. This dominance is reproduced by our fit, with the $[\rho\pi]\,{:}\,2^{+}D_{1}^{(-)}$ wave identified as the leading one in the pool of extracted waves. We emphasize that $2^{+}D_{1}^{(-)}$ is exactly the wave, in which the $a_2(1320)$ was mainly observed in the $\eta\,\pi^{-}$ mode~\cite{Albrecht:2024qdh}, as can be seen in the right panel of Figure~\ref{fig:a2-comparison}. 
%and $[\rho\pi]\,{:}\,2^{+}D_{1}^{(+)}$
%and $2^{+}D_{1}^{(+)}$

\section{Summary}
In these proceedings, we present the first stages of the amplitude analysis of a three-pion final state with GlueX-I data collected in 2017--2018. The analyzed final state is reconstructed in the $\gamma\,p\to \pi^{+}\pi^{-}\pi^{-}\Delta^{++}$ reaction with a linearly polarized beam. The objective of this analysis is to establish the existence of spin-exotic hybrid meson candidates, in particular the $\pi_{1}(1600)$, decaying into the three-pion ﬁnal state.

Preliminary results from our partial-wave fit to data using polarized amplitudes show a consistent picture of $a_{2}^{-}(1320)$ production off $\Delta^{++}$, in agreement with the GlueX amplitude analysis of the $\eta\,\pi^{-}$ mode~\cite{Albrecht:2024qdh}. Of particular interest is consistency in the prevalence of the negative reflectivity amplitudes and dominance of the spin projection $m=1$. In the course of subsequent work, intensities across $m_{3\pi}$  will be extracted for each amplitude in the fit model, with a dedicated study of the role of the spin-exotic $[\rho\pi]\,{:}\,1^{-}P$ wave. The analyzed $-t$ range will be extended to higher values. GlueX-I data will be further extended with new data collected in 2020--2027, thereby improving sensitivity to a potential $\pi_{1}(1600)$ signal.

\section{Acknowledgements}
This work was supported by the German Research Foundation (DFG) under project number 537210888. 
Calculations were partially performed on the HPC cluster Elysium of the Ruhr University Bochum, subsidised by the DFG (INST 213/1055-1). GlueX acknowledges the support of several sponsors (http://gluex.org/thanks).

\vspace*{-2mm}
\bibliography{MESON2026_IliaBelov_refs}

\end{document}